\documentclass[preprint2]{aastex6}
\usepackage{epstopdf}
\usepackage{lineno}

\shorttitle{Parameterizing the black hole's outflow in dSphs}
\shortauthors{Lanfranchi et al.}

\begin{document}

\title{Parameterizing the Outflow from a Central Black Hole in Dwarf Spheroidal Galaxies: a 3D Hydrodynamic Simulation}

\author{Gustavo A. Lanfranchi\altaffilmark{1}, Roberto Hazenfratz\altaffilmark{1}, Anderson Caproni\altaffilmark{1}, and Joseph Silk\altaffilmark{2,3,4}}

\altaffiltext{1}{N\'ucleo de Astrof\'\i sica, Universidade Cidade de S\~ao Paulo , R. Galv\~ao Bueno 868, Liberdade, 01506-000, S\~ao Paulo, SP, Brazil}
\altaffiltext{2}{Institut d’Astrophysique de Paris, UMR7095:CNRS \& UPMC, Sorbonne University, F-75014, Paris, France}
\altaffiltext{3}{Department of Physics and Astronomy, The Johns Hopkins University Homewood Campus, Baltimore, MD 21218, USA}
\altaffiltext{4}{Beecroft Institute of Particle Astrophysics and Cosmology, Department of Physics, University of Oxford, Keble Road, Oxford OX1 3RH, UK}


\begin{abstract}
Large galaxies harbor massive central black holes and  their feedback causes a substantial impact in their evolution. Recently, observations suggested that dwarf galaxies might host black holes in their centers, but with lower masses (intermediate-mass black holes - IMBH). The impact of such IMBHs on the evolution of the dwarf spheroidal galaxies (dSphs), however, has not been so far properly analysed. In this work, we investigate the effects of an outflow from an IMBH on the gas dynamics in dSph galaxies by means of non-cosmological, three-dimensional hydrodynamic simulations, letting the galactic gas distribution evolve over 3 Gyr under the influence of the IMBH's outflow and supernova feedback. All simulations have a numerical resolution of 20.0 pc cell$^{-1}$. Two scenarios are considered to infer the differences in the propagation of the outflow, one with a homogeneous ISM and another one with inhomogeneities caused by supernovae feedback. A minimal initial speed and a minimal initial density are required for the outflow to propagate, with the values depending on the conditions of the medium. In an
unperturbed medium, the outflow propagates freely in both directions with the same velocity (lower than the initial one), removing a small fraction of the gas from the galaxy (the exact fraction depends on the initial physical conditions of the outflow). However, in an inhomogeneous ISM, the impact of the outflow is substantially reduced, and its contribution to the removal of gas from the galaxy is almost negligible. 

\end{abstract}

\keywords{galaxies: dwarf --- galaxies: evolution --- galaxies: individual (Ursa Minor) --- galaxies: ISM --- hydrodynamics --- methods: numerical}

\section{Introduction} \label{sec:intro}

In the context of formation and evolution of galaxies, the study of dwarf galaxies is a particular issue which yields constraints on the first stages of galactic evolution in the young Universe and on the formation of larger systems at lower redshifts (\citealt{nfw95,moore99,robert05}). The Local Group is regarded as a key laboratory for such studies due to the morphological diversity of its galaxies and considering that it hosts the nearest dwarf galaxies. This proximity facilitates the observation and the detailed description of many chemical and physical properties of the stars and interstellar medium of many systems, which are necessary for computational simulations and for the development of evolutionary theories (\citealt{mate98,kuos00,hvg01,dolp02,ggh03,tht09,debo12a,debo12b}). 
	
The dwarf spheroidal galaxies (dSph) in the local Universe are fundamental objects in the study of galactic evolution. They are considered to be  independent remnants of the first small galaxies that merged gravitationally at higher redshifts to form larger and massive galaxies, in the context of the hierarchical models for structure formation under the $\Lambda$CDM framework (e.g. \citealt{robert05} and references therein for the Milky Way case). They are characterized by a predominant old stellar population in most cases, with no ongoing star formation and dominated by dark matter (\citealt{mate98}; \citealt{grceput09}). In this way, the observation and study of such kinds of galaxies would make it possible to analyze the advanced stages of such primordial galactic blocks (\citealt{nfw95}; \citealt{moore99}; \citealt{robert05}).  
	
One distinct feature of the dSphs of the Local Group is that practically no detectable HI regions were identified in any object, with all of them presenting 
$M_\mathrm{HI}/M_\mathrm{tot}$ $<$ 10$^{-3}$, where $M_\mathrm{HI}$ and $M_\mathrm{tot}$ are respectively the HI and total masses of the galaxies (\citealt{young00,ggh03}; \citealt{grceput09,spek14}). A fundamental question that rises from this is the identification of the physical mechanisms responsible for the exhaustion of their gas content (\citealt{mate98,grceput09}). In this regard, the relative importance of internal and external mechanisms, represented by ram pressure and tidal effects in the latter case and mainly supernova explosions in the former one (\citealt{mcfer99,grceput09,revjab12,gatto13,ruiz13,cap15,cap17}), is still a matter of debate.
	
\citet{nicbla11} argued that the dependence of the residual gas mass in local dwarfs with respect to the distance to the Galactic center can be considered as evidence of gas depletion caused by tidal effects and/or ram pressure. The occurrence of these processes are facilitated by high eccentricity in the orbits of  many dwarfs, whose gas could be removed by a hot halo around the Galaxy (\citealt{grceput09}).
On the other hand, \citet{ruiz13}, \citet{cap15}, \citet{cap17}, among others, argued that galactic winds triggered by supernovae (SNe) can be responsible for the removal of a large fraction of the gas of dwarf spheroidal galaxies, depending on the star formation rate considered. Another internal mechanism,  not yet fully explored, that can contribute to gas removal, is the outflow from a central black hole in these galaxies.

By extrapolating the relationship between the mass of the supermassive black holes (SMBH) and the mass of the host galaxy bulge (M$_\mathrm{BH}$-M$_\mathrm{Bulge}$) (\citealt{greene12}) down to the masses of intermediate-mass black holes (IMBH), one would expect the existence of such objects in the centers of systems with masses similar to  dwarf galaxies and globular clusters (\citealt{femer00}; \citealt{gult09}; \citealt{schu19}).

Over the past decade, several works have been published with analysis of observational evidence for intermediate-mass black holes or even supermassive ones ($M_\mathrm{BH}>10^6$ M$_\sun$) in the central regions of dwarf galaxies (\citealt{macc05}; \citealt{lora09}; \citealt{rein11}; \citealt{rein13}; \citealt{nuci13}; \citealt{mann15}; \citealt{mezcua20} and references therein). For the Ursa Minor dwarf galaxy, for example, \citet{macc05} estimated that the potential IMBH would have a mass of $\sim10^5$ M$_\sun$ and an X-ray luminosity of 10$^{34}$ erg s$^{-1}.$  \citet{lora09} analysed a double peak in the excess brightness density in the same galaxy by means of N-body simulations. They found an upper limit for an IMBH of (2 - 3) $\times$ 10$^4$ M$_\sun$ near the galactic center, which is consistent with the values for such objects expected by the extrapolation of the $M_\mathrm{BH}-\sigma$ relationship for elliptical galaxies (\citealt{gult09}). \citet{nuci13} and \citet{mann15} found spectroscopic evidences of coincident X-ray and radio compact sources, indicating a low-efficiency accreting black hole.
	
As in the case of Active Galactic Nucleus (AGN) counterparts powered by accreting SMBH, an IMBH in a dwarf galaxy might generate outflows that could affect both the evolution of its own black hole and its host galaxy due to its effect on the interstellar medium (ISM). Such outflows could also trigger or quench star formation depending on the nature of the ISM, as e.g., its density and degree of clumpiness. The two main mechanisms to explain  AGN feedback in a galaxy are the mechanical and radiative feedback that may inject momentum and thermal energy in the host galaxy (\citealt{moetal10}; \citealt{fgq12}; \citealt{zub14}). The mechanical or kinetic feedback is mediated by ram and thermal pressures, ignoring the radiation pressure. When only this feedback is considered, it is assumed that the radiative losses are low enough not to influence the outflow dynamics from the central compact object (\citealt{wagn12}). It may be argued that the mechanical outflow is dominant in galaxies where the accretion rate by the black hole is much smaller than the Eddington limit (\citealt{moetal10}). 
	
To the current knowledge of the authors, this paper presents the first hydrodynamic simulations of mechanical feedback in the form of an outflow from a possible IMBH in a dwarf spheroidal-like galaxy, in order to study its effect on the gas dynamics. The dSph galaxy chosen for the parameterization of the simulations is Ursa Minor, for which it was tested whether such an outflow could be responsible for any additional mass loss when compared to simulations where only supernova feedback is considered. 

Ursa Minor was discovered by Wilson in 1955 (\citealt{wils55}). Its heliocentric distance was estimated as 64 kpc (\citealt{irhat95}). Its stellar spatial distribution is best fit by a King profile with a 300 pc core radius and it has a tidal radius estimated as $\sim$ 0.9 - 1.5 kpc (\citealt{strig07}; \citealt{kirb11}). The velocity dispersion was estimated as 12 km s$^{-1}$ inside a radius of 36 arcsec from the nucleus (\citealt{wilk04}). It has a total mass of (5.3 $\pm$ 1.3) $\times$ 10$^7$ M$_\sun$ inside a 600 pc radius and mass-luminosity ratio $M/L\sim800$ (\citealt{strig07}; \citealt{strig08}; \citealt{wolf10}). The Ursa Minor dSph was chosen to simulate the effect of an IMBH outflow combined with supernovae due to the availability of data and previous studies regarding its chemical evolution (\citealt{lanfm03}; \citealt{lanfm04}; \citealt{lanfm07}) and three-dimensional (3D) hydrodynamic simulations (\citealt{cap15}; \citealt{cap17}), which analyzed the effect of supernovae as progenitors of galactic winds in the galaxy. Whereas \citet{lanfm03}, \citet{lanfm04} and \citet{lanfm07} were able to reproduce the chemical properties of UMi with a chemical evolution model characterized by a single star formation episode, which started $\sim$ 13 Gyr ago and lasted for 3 Gyr, and the occurrence of high-efficiency galactic winds caused by supernovae,	\citet{cap15} and \citet{cap17} suggested that  internal supernovae feedback alone is not enough to completely remove the gas out of the galaxy inside a radius of 600 pc, where most of the Ursa Minor stars reside (\citealt{irhat95}). These authors found that most of the gas can be removed, up to a gas mass of $\sim80\%$ with a time-varying supernovae rate.

The numerical setup and the initial conditions adopted in the hydrodynamic code are described in Section 2. In Section 3 the results are presented, followed by a discussion of the main results and comparison to previous works. The main conclusions of this work are summarized in Section 4.

\begin{table*} [t]
\begin{center}
\caption{Initial values for the velocity and density of the BH outflow. In the last columns, inferred values for the energy and luminosity of the outflow are given.}
\label{table1}
\begin{tabular}{l|ccccc}
\noalign{\smallskip}
\hline
\hline
\noalign{\smallskip}
Simulation & $n_\mathrm{out}$ & $v_\mathrm{out}$ & $u_\mathrm{kin}$ & $u_\mathrm{th}/u_\mathrm{kin}$ & $L_\mathrm{kin}$\\
& (cm$^{-3}$) & (km s$^{-1}$) & (erg cm$^{-3}$) &  & (erg s$^{-1}$)\\
\noalign{\smallskip}
\hline
\noalign{\smallskip}
JET001100 & 0.001 & 100 & 8.30 $\times 10^{-14}$ & 1.206 & 6.97 $\times 10^{33}$\\
JET0011000 & 0.001 & 1000 & 8.30 $\times 10^{-12}$ & 0.012 & 3.20 $\times 10^{36}$\\
JET0012000 & 0.001 & 2000 & 3.32 $\times 10^{-11}$ & 0.003 & 2.54 $\times 10^{37}$\\
JET0013000 & 0.001 & 3000 & 7.47 $\times 10^{-11}$ & 0.001 & 8.55 $\times 10^{37}$\\
JET0031000 & 0.003 & 1000 & 2.49 $\times 10^{-11}$ & 0.012 & 9.60 $\times 10^{36}$\\
JET0051000 & 0.005 & 1000 & 4.15 $\times 10^{-11}$ & 0.012 & 1.60 $\times 10^{37}$\\
JET0052000 & 0.005 & 2000 & 1.66 $\times 10^{-10}$ & 0.003 & 1.27 $\times 10^{38}$\\
JET011000 & 0.01 & 1000 & 8.30 $\times 10^{-11}$ & 0.012 & 3.20 $\times 10^{37}$\\
JETSNE001100 & 0.001 & 100 & 8.30 $\times 10^{-14}$ & 1.206 & 6.97 $\times 10^{33}$\\
JETSNE0031000 & 0.003 & 1000 & 2.49 $\times 10^{-11}$ & 0.012 & 9.60 $\times 10^{36}$\\
JETSNE0051000 & 0.005 & 1000 & 4.15 $\times 10^{-11}$ & 0.012 & 1.60 $\times 10^{37}$\\
JETSNE0052000 & 0.005 & 2000 & 1.60 $\times 10^{-10}$ & 0.003 & 1.27 $\times 10^{38}$\\
JETSNE011000 & 0.01 & 1000 & 8.30 $\times 10^{-11}$ & 0.012 & 3.20 $\times 10^{37}$\\
\hline
\end{tabular}
\end{center}
\end{table*}

\section{Numerical setup and initial conditions} \label{sec:NumSetup}

Non-cosmological, 3D hydrodynamic simulations of the gas of a typical isolated dSph galaxy are performed to analyse the role played by the outflow of a central intermediate-mass black hole in its internal dynamics and in  gas removal from  the system. Assuming an initial baryonic-to-dark-matter ratio inferred from the cosmic microwave background radiation and a cored and static dark matter gravitational potential, the galactic gas distribution is evolved for 3 Gyr taking into account the outflow of a central IMBH and SNe feedback. The initial setup of the simulation is exactly the same as one adopted for Ursa Minor dSph (used as a template for a classical dSph galaxy), described in detail in \citet{cap15} and \citet{cap17}. The interstellar medium is initially in hydrostatic equilibrium with the dark matter potential (as suggested by \citealt{mcfer99}). The total mass of the dark matter halo is estimated from the core radius $r_0$ = 300 pc, the initial sound speed in the medium c$_{s0}$ = 11.5 km s$^{-1}$ and the circular velocity inferred for Ursa Minor $v_\mathrm{c}$ = 21.1 km s$^{-1}$ (\citealt{irhat95}; \citealt{strig07} ), yielding a value of $M_\mathrm{h} \sim 1.51 \times 10^9$ M$_\sun$. The estimated initial total gas mass is $M_\mathrm{g0} \sim 2.94 \times 10^8$ M$_\sun$ and the initial central density is $\rho_0 \sim 4.6 \times 10^{-23}$ g cm$^{-3}$. The galaxy is simulated inside a computational cube with sides of 3 kpc length, each divided in 150 cells yielding a resolution of {\bf 20} pc cell$^{-1}$. Our simulation comprises 3 Gyr, the same duration estimated for the star formation in Ursa Minor, and was performed in the supercomputer SDumont\footnote{Further information on SDumont cluster at \url{http://sdumont.lncc.br}.}  at the National Laboratory for Scientific Computing (LNCC/MCTI, Brazil) and using the message passing interface (MPI) library for parallelization.

The outflow from the IMBH leaves the central region of the galaxy
along the $z-$axis of a Cartesian coordinate system ($x,y,z$) in both directions.
It is simulated with the injection of a fluid with number density $n_\mathrm{out}$, temperature $T_\mathrm{out}$, and speed $v_\mathrm{out}$
in the central cell of the computational cube ($x = y = z = 0$). Different values for $n_\mathrm{out}$ and $v_\mathrm{out}$ are adopted in each simulation, but the outflow temperature in the inlet region is kept the same: $T_\mathrm{out} = 2.9\times 10^5$ K.

 \begin{figure*}
  \centering
  \includegraphics[width=18cm]{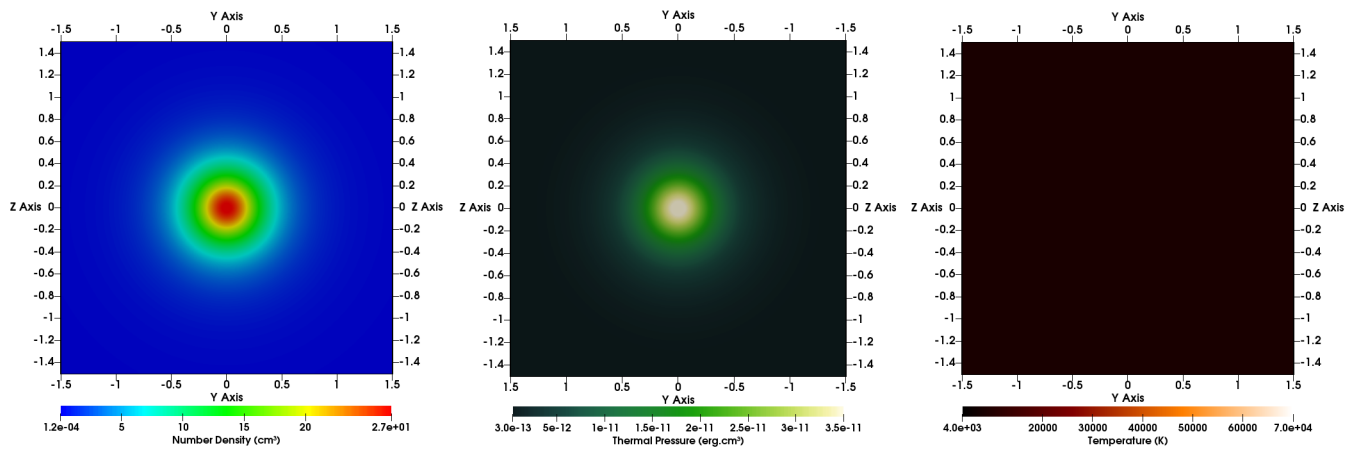}
 \caption{Cut in the $x = 0$ plane of the galaxy at $t = 0$ yr, showing the gas number density (left panel), pressure (middle panel) and temperature (right panel) profiles.}
 \label{initial cut}
 \end{figure*}

\subsection{The hydrodynamic code} \label{sec:code}

The classical hydrodynamic differential equations are solved by the numerical code PLUTO\footnote{\url{http://plutocode.ph.unito.it/}} \citep{mig07}:

\begin{equation} \label{massconserv}
\frac{\partial \rho}{\partial t} + \nabla\cdot\left(\rho\textbf{\emph{v}}\right) = 0,
\end{equation}

\begin{equation} \label{momconserv}
\frac{\partial \left(\rho\textbf{\emph{v}}\right)}{\partial t} + \nabla\cdot\left(\rho\textbf{\emph{v}}\textbf{\emph{v}} + P\mathbf{I}\right) = -\rho\nabla\Phi,
\end{equation}

\begin{equation} \label{energconserv}
\frac{\partial E}{\partial t} + \nabla\cdot\left[\left(E+P\right)\textbf{\emph{v}}\right] = F_\mathrm{c} - \rho\textbf{\emph{v}}\cdot\nabla\Phi,
\end{equation}
\\where the mass density is represented by $\rho$, $\textbf{\emph{v}}=(v_\mathrm{x}, v_\mathrm{y}, v_\mathrm{z})^T$ is the fluid velocity in Cartesian coordinates, the thermal pressure is $P$,  $\mathbf{I}$ is the identity tensor of rank 3, and the total energy density is given by $E$

\begin{equation} \label{totE}
E = \rho\epsilon + \frac{\rho\vert\textbf{\emph{v}}\vert^2}{2},
\end{equation}
\\where $\rho\epsilon$ is the internal density energy. The ideal equation of state $P = (\Gamma-1)\rho\epsilon$ was adopted in this work, where $\Gamma$ is the adiabatic index of the plasma, assumed as 5/3. This choice implies in a sound speed of the plasma, $c_\mathrm{s}$, defined as $c_\mathrm{s}=\sqrt{\Gamma P/\rho}$.

The quantity $F_\mathrm{c}$ in equation (\ref{energconserv}) is the cooling function

\begin{equation} \label{cooling}
F_\mathrm{c}=\frac{\partial P}{\partial t} = - \left(\Gamma-1\right)n^2\Lambda(T),
\end{equation}
where $n$ is the number density of the gas and $\Lambda(T)$, obtained from the interpolation of the precomputed tables provided by \citet{wie09}. To compute the values of $\Lambda(T)$, we follow \citet{cap15}, adopting $n=0.1$ cm$^{-3}$, roughly the initial mean number density inside the simulated box domain, and [Fe/H]$\sim-2.13$, the median metallicity of \object{Ursa Minor} \citep{kirb11}.

We used in our simulation the same initial gas density and pressure profiles of the models Mgh19SN1 and Mgh19SN10 found in \citet{cap15}. They were fully determined imposing hydrostatic equilibrium between an initial, isothermal gas with sound speed, $c_\mathrm{s_0}$ = 11.5 km s$^{-1}$, and a cored, static dark matter (DM) gravitational potential, $\Phi_\mathrm{h}$ \citep{bitr87,mcfer99}: 

\begin{equation} \label{dm_pot} 
\Phi_\mathrm{h}(\xi) = v^2_\mathrm{c_\infty}\left[\frac{1}{2}\ln(1+\xi^2)+\frac{\arctan\xi}{\xi}\right],
\end{equation}
generated by an isothermal, spherically symmetric DM mass density profile with a characteristic radius, $r_0$, equals to 300 pc. In equation (\ref{dm_pot}), $\xi = r/r_0$, $r$ is the radial distance, and $v_\mathrm{c_\infty}$ is the maximum circular velocity due to the DM gravitational potential, assumed to be equal to 21.1 km s$^{-1}$ (in agreement with \citealt{strig07}).

\begin{figure*}
  \centering
  \includegraphics[width=18cm]{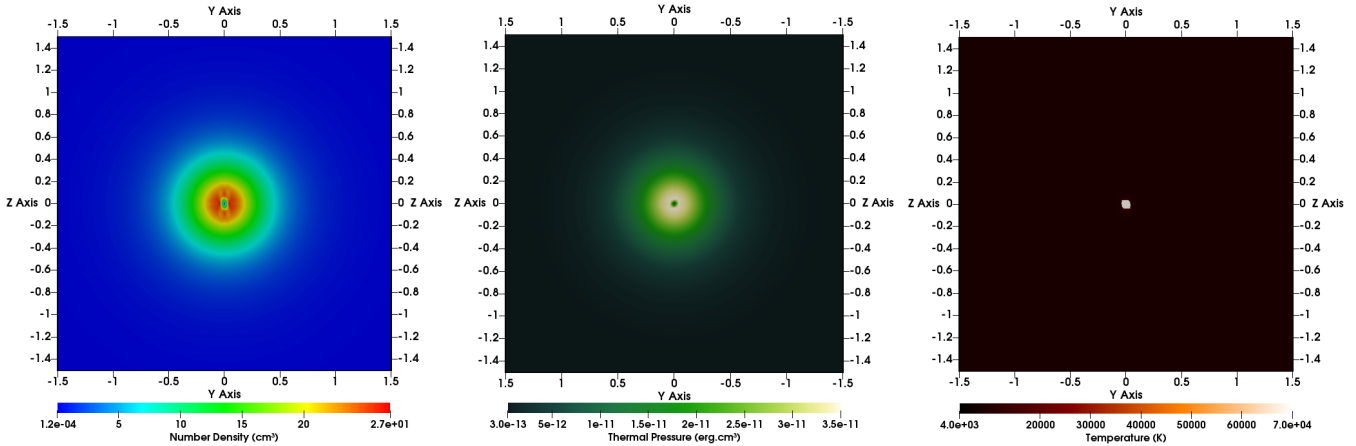}
\caption{Cut in the x = 0 plane of the galaxy at t = 125 Myr, showing the gas number density (left panel), pressure (middle panel) and temperature (right panel) profiles for simulation JET0011000.}
\label{no_outflow}
\end{figure*}

\section{Results} \label{sec:results}

\subsection{The Parameter Space} \label{subsec:spaceparameter}

In our simulations, the outflow is characterized by the initial values adopted for its density ($n_\mathrm{out}$) and speed ($v_\mathrm{out}$), two free parameters that are constrained indirectly by observations, as well as by the temperature. The initial density of the outflow is constrained by the range of masses observationally estimated for an IMBH in the center of the dwarf spheroidal galaxy Ursa Minor: $M_\mathrm{BH}\sim 10^4 - 10^6$ M$_{\sun}$ (\citealt{macc05}; \citealt{lora09}; \citealt{nuci13}; \citealt{mann15}), whereas the initial speed is limited by speeds inferred from molecular lines detected in outflows associated with AGNs in larger galaxies, theoretical works (\citealt{cuieta20}) and in outflows driven by AGNs in dwarf galaxies (\citealt{manza2019}; \citealt{liu20}). 
The temperature of the outflow is the same in all scenarios ($T_\mathrm{out}$ = 2.9 $\times$ 10$^5$ K), whereas the speed and the density initial values are varied separately and the effects of the BH's outflow in the density, radial velocity, and thermal pressure of the galactic gas are analysed in different scenarios, in order to define a range of values that allows it to cause some discernible impact in the ISM of a classical dwarf spheroidal galaxy.

Two scenarios for the ISM of the galaxy are considered for the parameter space study: an almost homogeneous medium with the gas density varying only with the galactic radius, and a medium with local inhomogeneities caused by explosions of supernovae. In both scenarios the outflow is switched on at $t = 0$ yr, when the gas is in hydrostatic equilibrium with the dark matter potential of the galaxy (Figure \ref{initial cut}), and turned off after 250 Myr. There is not a consensus regarding the time-scale of the AGN phase, but indirect estimates lie inside a broad range from 10$^7$ to 10$^9$ yr (\citealt{solt82}; \citealt{mart01}; \citealt{yutr02}; \citealt{marc04}). During this time-scale, the AGN activity could not be continuous, as pointed out by \citet{scha15} who presented observational evidence that AGNs could turn on and off in intervals of 10$^5$ yr. In the case of an inhomogeneous medium, the hydrostatic equilibrium of the ISM is disturbed by the SNe, but all other assumptions remain the same. The energy of each supernova (SN) explosion is injected in the medium following the prescriptions of \citet{cap15} and \citet{cap17}: an energy of 10$^{51}$ erg is added in a volume defined by a specific number of cells ({\bf one cell} in this work). The cells selected as  possible sites for the injection of energy are randomly chosen, weighted by the gas density in each cell (the higher the density of the cell, the higher is the probability of the energy being injected there). The distribution of the SNe explosions in time is given by the rate of SNe taken from chemical evolution models which take into account the stellar lifetimes and reproduces the observed chemical properties of the galaxy \citep{lanfm07}. Since the stellar lifetimes are properly considered and the outflow starts at $t = 0$ yr, the SNe begin to disturb the medium a few Myr after the outflow is already set.

 \begin{figure*}
 \centering
 \includegraphics[width=13cm]{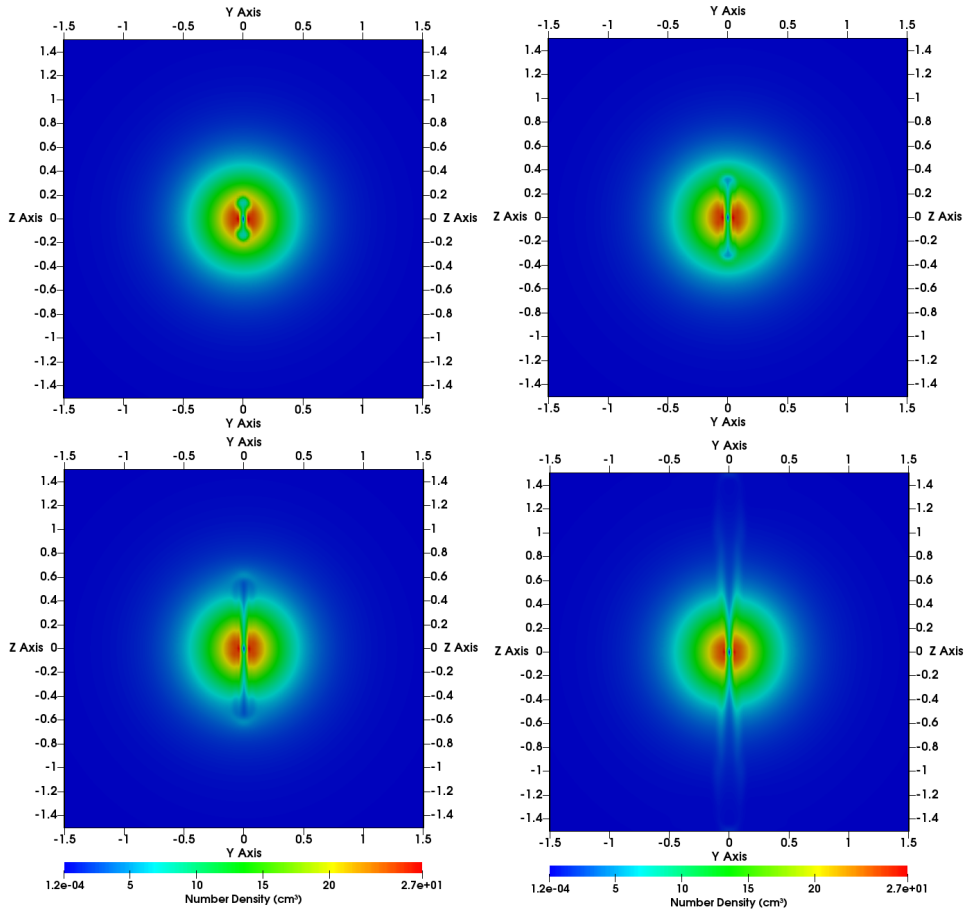}
 \caption{Cut in the $x = 0$ plane for the gas density of the gas at t = 25 Myr (upper left), 50 Myr (upper right), 75 Myr (bottom left), and 150 Myr (bottom right) for the simulation JET0012000.}
 \label{2000vel}
 \end{figure*}

Even though the scenario with an almost homogeneous medium with the gas density varying only with the galactic radius constitutes a simple assumption, it helps us to identify the importance of each parameter of the outflow on the ISM dynamics. We begin the analysis by testing different values for the initial speed and the initial density in an ``homogeneous" medium followed by the scenario with the disturbed medium. The main physical parameters of the outflows simulated in this work are listed in Table \ref{table1}. Besides $n_\mathrm{out}$ and $v_\mathrm{out}$, we also provided the values of the kinetic energy density, $u_\mathrm{kin}$, defined as

\begin{equation} \label{ukin}
u_\mathrm{kin} = \frac{1}{2}\rho_\mathrm{out} v_\mathrm{out}^2,
\end{equation}
\\where $\rho_\mathrm{out}$ is the mass density of the outflow at its inlet region, the ratio between $u_\mathrm{kin}$ and the thermal energy density, $u_\mathrm{th}$, 

\begin{equation} \label{uth}
u_\mathrm{th} = \frac{\Gamma}{\Gamma-1}P_\mathrm{out},
\end{equation}
\\where $P_\mathrm{out}$ is the thermodynamic pressure of the outflow at its inlet region, and the kinetic luminosity, $L_\mathrm{kin}$, which is defined in the non-relativistic regime as (e.g., \citealt{per17})

\begin{equation} \label{Lkin}
L_\mathrm{kin} = \frac{1}{2}\rho_\mathrm{out} v_\mathrm{out}^3 A_\mathrm{out}\left(1+\frac{u_\mathrm{th}}{u_\mathrm{kin}}\right),
\end{equation}
\\where $A_\mathrm{out}$ is the outflow cross-section\footnote{Area of the square faces of the central computational cell where the outflow is injected.}.

Our results show that only outflows with initial speeds similar or higher than 1000 km s$^{-1}$ and number densities similar or higher than 0.003 cm$^{-3}$ impact the ISM of the galaxy, whether or not the ISM is quasi-homogeneous. In this case, a thin jet structure is formed along the direction 
of the initial velocity. In scenarios with $v_\mathrm{out} < 1000$ km s$^{-1}$ and $n_\mathrm{out} <$ 0.003 cm$^{-3}$ (Simulations JET001100, JET0011000, JETSNE001100, JETSNE0011000), the propagation of the outflow is not visible (no jet structure is formed) in the profile of the gas density, pressure or temperature as it is shown in Figure \ref{no_outflow}. In an homogeneous ISM, the gas density is higher at the center of the galaxy ($\rho_0$ = 4.6 $\times$ 10$^{-23}$ g cm$^{-3}$) preventing outflows with lower initial density and speed to propagate, as they do not have enough energy to leave the central cell. If the SNe feedback is taken into account, it lowers the gas density of the central region of the galaxy after a few SNe explosions. But at the same time these SNe are responsible for increasing the pressure of the surrounding gas, which makes difficult (or even impossible) for the outflow to develop. As $n_\mathrm{out}$ is increased, the outflow gets more energetic and becomes able to propagate across the ISM, whether or not it is homogeneous. The higher is $n_\mathrm{out}$, the easier it is for the outflow to leave the central region of the galaxy and propagate through the ISM, with also higher propagation velocities.

\begin{figure*}
 \centering
 \includegraphics[width=13cm]{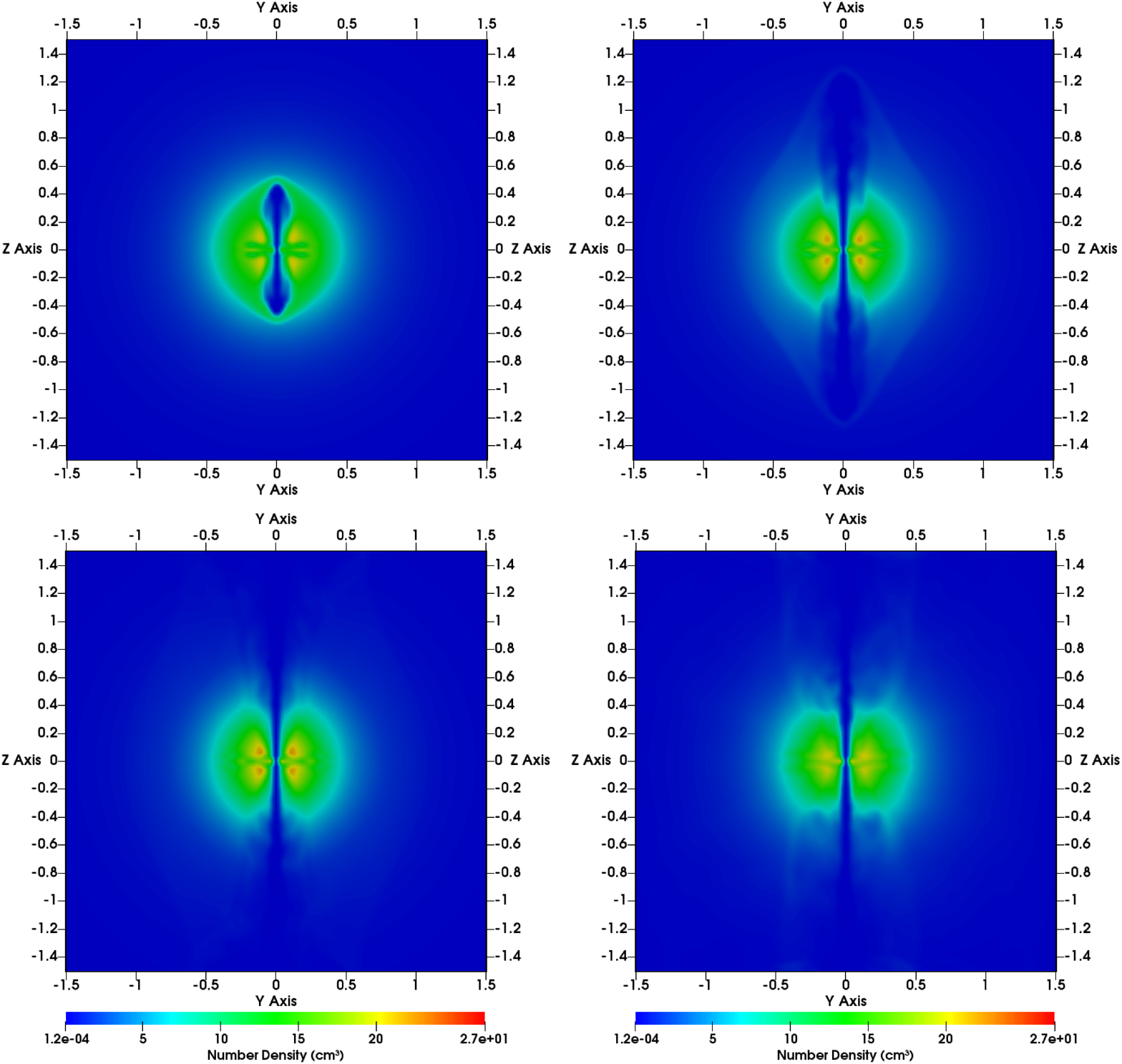}
\caption{Cut in the $x = 0$ plane for the density of the gas at $t = 25$ Myr (upper left), 50 Myr (upper right), 75 Myr (bottom left), and 150 Myr (bottom right) for the simulation JET0013000.}
\label{3000vel}
\end{figure*}

\begin{figure*}[t]
\centering
\includegraphics[width=17.7cm]{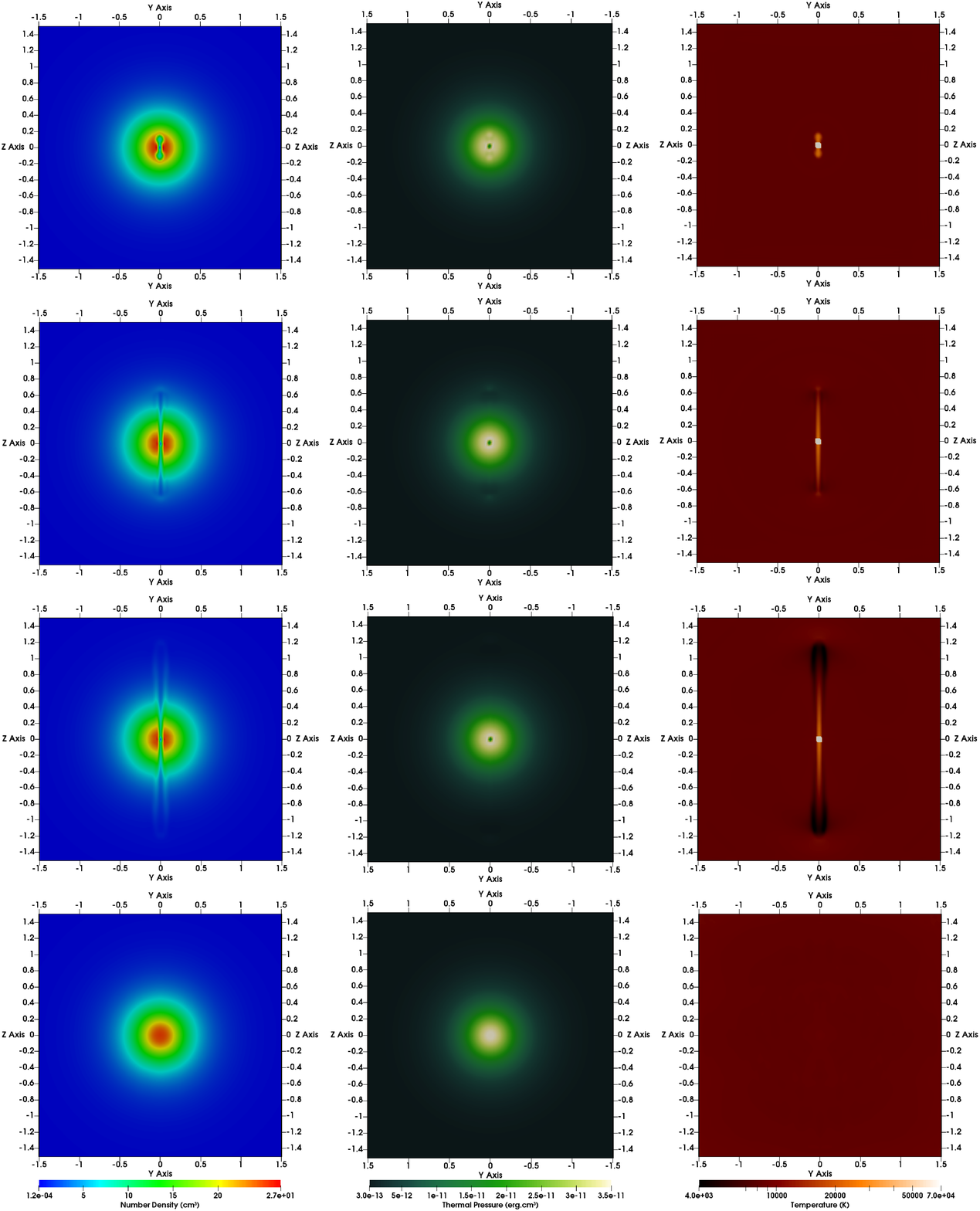}
\caption{Cut in the $x = 0$ plane of the galaxy at $t = 25$ Myr (upper line), $t = 100$ Myr (second line), $t = 150$ Myr (third line) and $t = 3$ Gyr (lower line), showing the gas density (left panel), thermal pressure (middle panel) and temperature (right panel) profiles for simulation JET0031000.}
\label{003density}
\end{figure*}

\begin{figure*}[t]
\centering
\includegraphics[width=17.7cm]{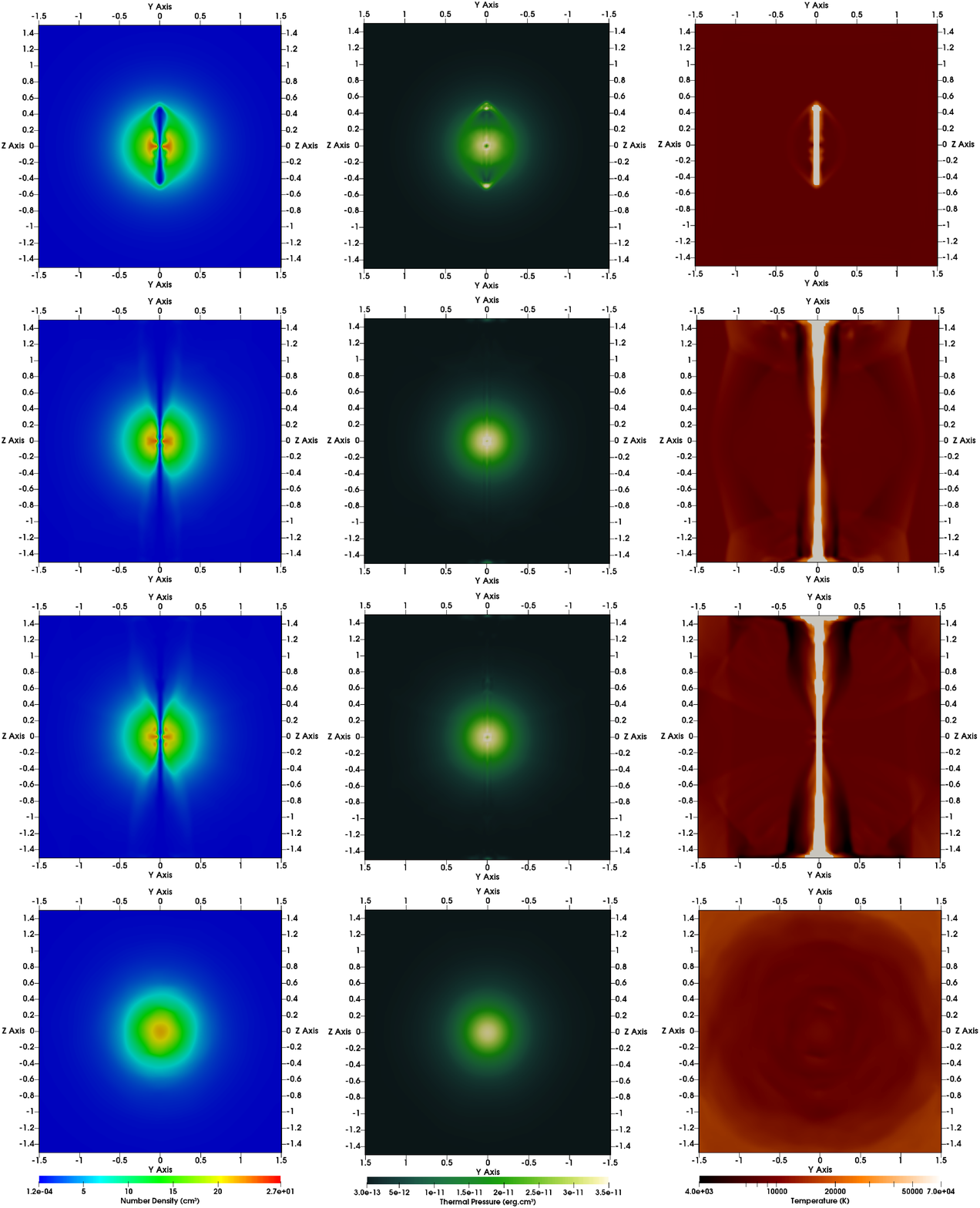}
\caption{Cut in the $x = 0$ plane of the galaxy at $t = 25$ Myr (upper line), $t = 100$ Myr (second line), $t = 150$ Myr (third line) and $t = 3$ Gyr (lower line), showing the gas density (left panel), thermal pressure (middle panel) and temperature (right panel) profiles for simulation JET011000.}
\label{01density}
\end{figure*}

On the other hand, changes in  $v_\mathrm{out}$ do not increase its impact on the ISM of the galaxy if $v_\mathrm{out}\la1000$ km s$^{-1}$. In such cases the energy of the outflow is too low to produce any impact on neighboring regions of the outflow inlet cell. For $v_\mathrm{out} > 1000$ km s$^{-1}$, however, the situation changes drastically in the case of an almost homogeneous medium. In the simulation JET0012000 with the outflow's lowest initial density ($n_\mathrm{out} = 0.001$ cm$^{-3}$) and $v_\mathrm{out}= 2000$ km s$^{-1}$, the effects of the outflow in an almost homogeneous galactic ISM are clear, as it is shown in Figure \ref{2000vel}. Initially, 25 Myr after the onset of the outflow, two bubble features can be noticed moving in the axis of the outflow's propagation, leaving the highest gas density region of the galaxy and pushing away gas of the interstellar medium. A lower density path (with one third of the initial central gas density and 100 pc wide) is created, making easier for the the outflow to build a thin jet feature leaving the central 400 pc radius of the galaxy (after 75 Myr) and reaching its outskirts far away from the tidal radius (1.5 kpc at 150 Myr). 

The overall effect on the ISM is much more significant when the outflow's initial speed is even higher ($v_\mathrm{out} = 3000$ km s$^{-1}$ - simulation JET0013000), as it can be seen in Figure \ref{3000vel}. Soon after the outflow begins, a broad jet structure (almost 250 pc wide) can be noticed propagating towards positive and negative {\it z} and pushing the galactic gas to the outer regions of the galaxy. After the onset of the outflow, the central region of the galaxy has its gas density decreased to half of its initial value. After 25 Myr, the gas speed in the $z$ direction at the shock front (at 400 pc) is around 1100 km s$^{-1}$. As the outflow propagates, it leaves the tidal radius of the galaxy before 50 Myr and the gas speed increases again to $\sim$ 1700 km s$^{-1}$ in the lowest density regions of the intergalactic medium. The gas leaves the computational domain (1.5 kpc) with this speed at the axis of the outflow. 

This high speed, with which the outflow reaches the intergalactic medium, is in the same order as those  estimated in galaxies hosting AGNs and similar to the speeds estimated by \citet{manza2019} and \citet{liu20}. \citealt{manza2019} detected and measured extended outflows in a sample of 13 dwarf galaxies with masses ranging from 3.63 $\times$ 10$^8$ to 9.33 $\times$ 10$^9$ M$_{\sun}$ and estimated outflow speeds between 375 and 1090 km s$^{-1}$, similar to the speeds from 240 km s$^{-1}$ to 1200 km s$^{-1}$ found by \citealt{liu20} through an integral-field spectroscopic study of eight dwarf galaxies. Hence  we decided to limit our parameter space for the initial density and initial speed of the outflow (the values that would allow it to impact the ISM of the galaxy) within the ranges: $n_\mathrm{out}$ = 0.003 -- 0.1 cm$^{-3}$ and $v_\mathrm{out}$ = 1000 -- 2000 km s$^{-1}$. In the next sections, we explore the interaction between outflows with these physical conditions and the ISM in a quasi-homogeneous phase and disturbed by SNe explosions. We also analyse the outflow influence on the gas removal of the galaxy calculating the remaining gas inside different radii as a function of time.

\subsection{The ``Homogeneous" Medium} \label{subsec:homo}

The gas density, pressure and temperature profiles in the $x = 0$ plane of the galaxy shown in Figures \ref{003density} and \ref{01density} reveal the dynamics of the interstellar gas due to the BH's outflow with different initial densities in the scenario where the gas density of the medium varies only with the radius. The impact of the outflows with different initial densities $n_\mathrm{out} = 0.003$ cm$^{-3}$, 0.005 cm$^{-3}$, and 0.01 cm$^{-3}$, but the same initial speed ($v_\mathrm{out}$ = 1000 km s$^{-1}$) in the ISM, are qualitatively similar. In all cases, a jet feature is created in the direction of the axis of the outflow's initial velocity (as seen in the second and third lines of the first column on
Figures \ref{003density} and \ref{01density}). This jet propagates almost freely and symmetrically in both directions of the axis and practically does not affect the ISM of the galaxy, unless near its propagation direction. As can also be noted in the number density and pressure panels shown in Figures \ref{003density} and \ref{01density}, jet interactions with the ISM produced usual bow shock structures seen in numerical simulations dealing with supersonic jets (e.g., \citealt{subi07,ros08,wabi11,wagn12,hakr13,wagn16,muk18}).
The speed of the gas inside the jet feature, in each case, is almost the same at every point, but below the initial one ($v_\mathrm{out}$). An outflow with a higher initial density, however (Simulation JET011000), gives the gas inside the jet feature a higher propagation speed throughout the medium. Around 50 Myr after the beginning of the simulation, the average speed of the gas inside the ``jet" is lower than $\sim$ 21 km s$^{-1}$ for Simulation JET0031000, $\sim$ 520 km s$^{-1}$ for Simulation JET0051000, and $\sim$ 730 km s$^{-1}$ for Simulation JET011000, all within the range estimated by \citet{manza2019,liu20}. Consequently, the outflow reaches the tidal radius of the galaxy (950 pc) sooner when its initial density is higher. One can see in the top panels in the left column of Figures \ref{003density} and \ref{01density} that at 25 Myr the outflow with higher initial density (Simulation JET011000) is breaking out the ISM region with higher gas density ($r\sim$ 500 pc) whereas the lower density outflow is still inside the central $\sim$ 200 pc of the galaxy. In the case of the Simulation JET0031000 (Figure \ref{003density}), one can see in the third line of the first column ($t = 150$ Myr) that the ISM is virtually undisturbed, even after the jet feature has reached out the external regions of the galaxy. Only regions close to the axis of the outflow propagation are disturbed. A thin jet feature is created moving through the ISM, leaving a region of low gas density and high temperature behind it. The gas is pushed away by the front of the jet feature until it leaves the tidal radius of the galaxy, spreading gas in the intergalactic medium.  

When the initial density of the outflow is higher ($n_\mathrm{out}$ = 0.01 cm$^{-3}$ - Simulation JET011000), on the other hand, the jet feature affects the propagation of the gas also in the medium surrounding it. In the gas density and temperature profiles (respectively left and right column of Figures \ref{003density} and \ref{01density}) the different effects of the outflows with different densities in the ISM can be clearly seen. As in the case of Simulations JET0031000 and JET0051000, a jet feature is created pushing the gas in its way out of the galaxy, leaving a region of low gas density behind. The difference in this case is that the "jet" is much thicker, affecting also the surrounding medium. The same effect can be seen in the temperature profile: a higher density outflow has a larger impact in the medium temperature. The initial medium temperature of the medium ($T_0$ $\sim$ 9.5 $\times$ 10$^3$ K) increases reaching almost 1.1 $\times$ 10$^4$ K after 225 Myr, with peaks of 2.3 $\times$ 10$^4$ K at points 100 pc from the jet feature.

\begin{figure}
\centering
\includegraphics[width=8cm]{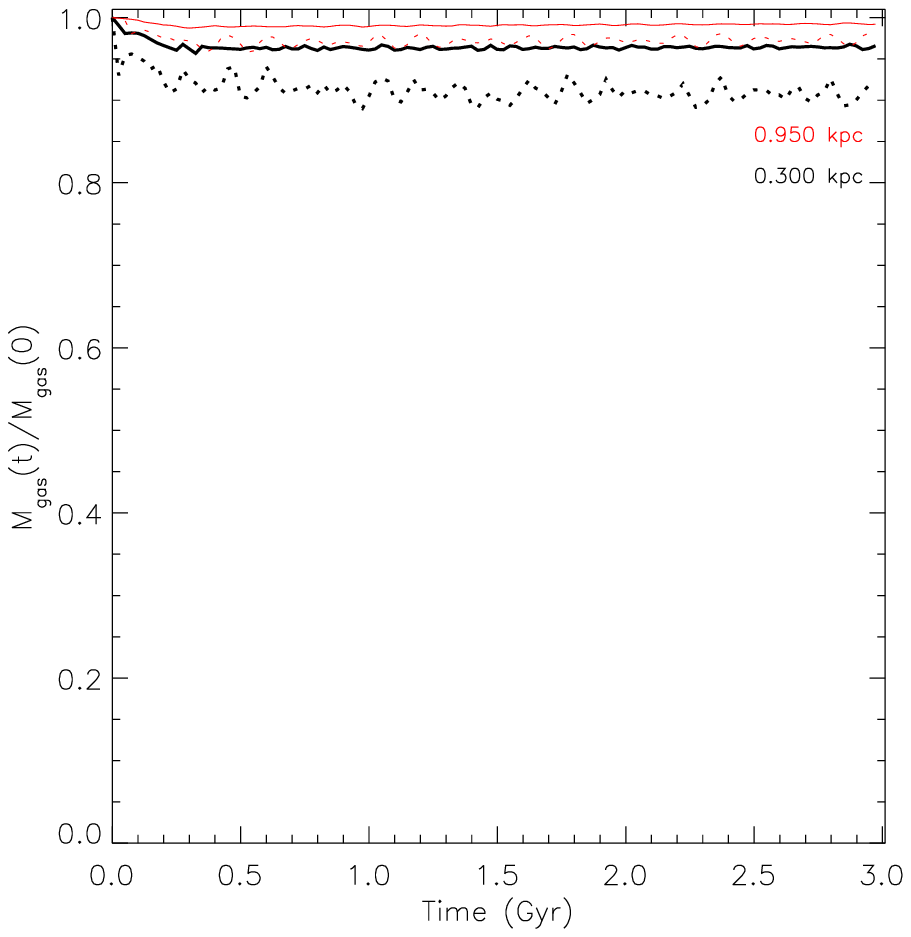}
\caption{Mass fraction as a function of time inside two different galactic regions (300 pc - black thick lines -  and 950 pc - red thin lines) for the Simulations JET0031000 (solid lines) and JET011000 (dashed lines).}
\label{gas_fraction_jet00301}
\end{figure}

In this simple scenario, we compute the mass of gas that is pushed away by the BH's outflow and leaves the galaxy. In Simulation JET0031000 only an insignificant fraction of the gas is removed from
the galaxy after 3 Gyr of simulation. The gas that is lost from the spherical region inside the tidal radius of the galaxy corresponds to almost $1\%$ of the initial gas mass (solid red thin line in Figure \ref{gas_fraction_jet00301}). Since this outflow creates a very thin jet feature along the axis of the propagation, only the gas in its path is carried and pushed away when the jet is developing. After this feature is established it creates an almost free path through which the outflow energy escapes the galaxy, without removing gas. On the other hand, in the case of Simulation JET011000, where the initial density of the outflow is higher, the fraction of gas that is lost increases. After the outflow leaves the central region of the galaxy the thicker jet feature pushes a larger fraction of gas. At the end of the simulation (after 3 Gyr) almost 4$\%$ of the initial gas was removed from the spherical region inside the tidal radius of system (solid black thick line in Figure \ref{gas_fraction_jet00301}). Even in this case, the removal of gas is meaningful only when the jet feature is being created. In both cases, the outflow is more effective inside the core radius of the galaxy (black thick lines in Figure \ref{gas_fraction_jet00301}), where it causes an impact in the ISM. Almost 10$\%$ of the initial gas is lost in the case JET011000, whereas in the case of a low density outflow (JET0031000), only 3$\%$ of the gas is removed from the galaxy. This result shows that the outflow alone, in the conditions simulated, can remove gas of the galaxy, but only a very small fraction and mainly from its central region. Several other factors must play a important role on the gas loss. A critical internal mechanism, the feedback of SNe, should also be taken into account.

\subsection{The Disturbed Medium} \label{subsec:disturbed}

In the simulations JETSNE0011000, JETSNE0031000, JETSNE0051000, JETSNE0052000 and JETSNE011000 the feedback of SNe is also considered simultaneously with the outflow from the IMBH. The number of the SNe that explode during  3 Gyr and the epoch of each supernova occurrence in the simulations are taken from the SNe rates of the chemical evolution model of \citet{lanfm07}, which reproduces several chemical properties of the dSph galaxy Ursa Minor. {Even though the gas mass and the dark halo mass adopted in the hydrodynamic simulations are different from the ones in the chemical evolution models (difference by factors of two and three, respectively), the predicted chemical properties by a model with these new masses are not significantly changed (only 0.2 to 03 dex in the [Fe/H], for instance) if the same SNe rates are achieved, and its results are still compatible with the observed chemical properties of the galaxy.}

Two important features of the Lanfranchi $\&$ Matteucci chemical evolution models are the stellar lifetimes, which are taken into account, and SNe types II and Ia that are properly considered. Following \citet{cap17}, we consider the sum of SNe Ia and SNe II rates without separating one type from another, since in both cases the same amount of energy is released in each explosion. There is, however, a change in the sites of explosion of the SNe when the rate of SNe Ia exceeds the rate of SNe II. The prescription for the spatial distribution of the energy injection is as follows: in the first 1.8 Gyr of the evolution of the galaxy (when the SNe II rate is predominant) the computational cells in which the energy of each SN (10$^{51}$ erg) is injected are randomly chosen, but weighted by the density of the gas. When the SNe rate indicates that a supernova must occur, the code chooses the cell with the highest gas density as a probable site for the injection of energy. If there is more than one cell with the highest gas density, one of them is randomly chosen. After $t \sim 1.8$ Gyr, when the SNe Ia rate becomes higher than the SNe II rate, SNe can occur anywhere: the choice of the cell where the energy will be injected becomes completely random. These prescriptions are justified by the timescales of each type of SN: type II SNe are characterized by a short timescale (almost 30 Myr for a progenitor of $\sim 8$ M$_\odot$) and might occur in the dense region where the star was formed, whereas SNe Ia can take a much longer time (up to billion years) to occur and can explode far from their parent molecular clouds. The choice of considering only SNe comes from the different amounts of energy released by these two phenomena: $\sim$ 10$^{51}$ erg in the case of supernovae and $\sim$ 10$^{35}$ erg by stellar winds (\citealt{brada98}, \citealt{roma19}). It is also justified by the results of \citet{roma19}, who performed high-resolution 3D simulations to examine the role played by stellar feedback in the Boötes I ultra faint dwarf galaxy considering the energy and mass injected in the IMS by OB associations and supernovae.  The effect of the energy from OB associations is negligible and the medium is hardly disturbed. Only after the onset of the SNe one can notice modifications in the ISM, even though not sufficient to remove gas from the galaxy.

The energy released by the SNe disturbs the medium, initially changing locally its gas density, pressure, and temperature. After a few tens of Myr, the almost homogeneous medium with the density varying only with the radius is drastically altered. Complex structures are created disrupting the spherically symmetric distribution of the gas density, due to multiple interactions of SNe remnants. Regions of outflow and inflow of gas can be seen, as well as regions of high density and high pressure side by side with regions of low density, pressure and temperature (see details in \citealt{cap15} and \citealt{cap17}). In such a medium, if the initial density or the initial speed of the outflow are not high enough (inside the range defined by the parameter space), its effect on the ISM is unnoticeable.

\begin{figure*}[t]
\centering
\includegraphics[width=17.8cm]{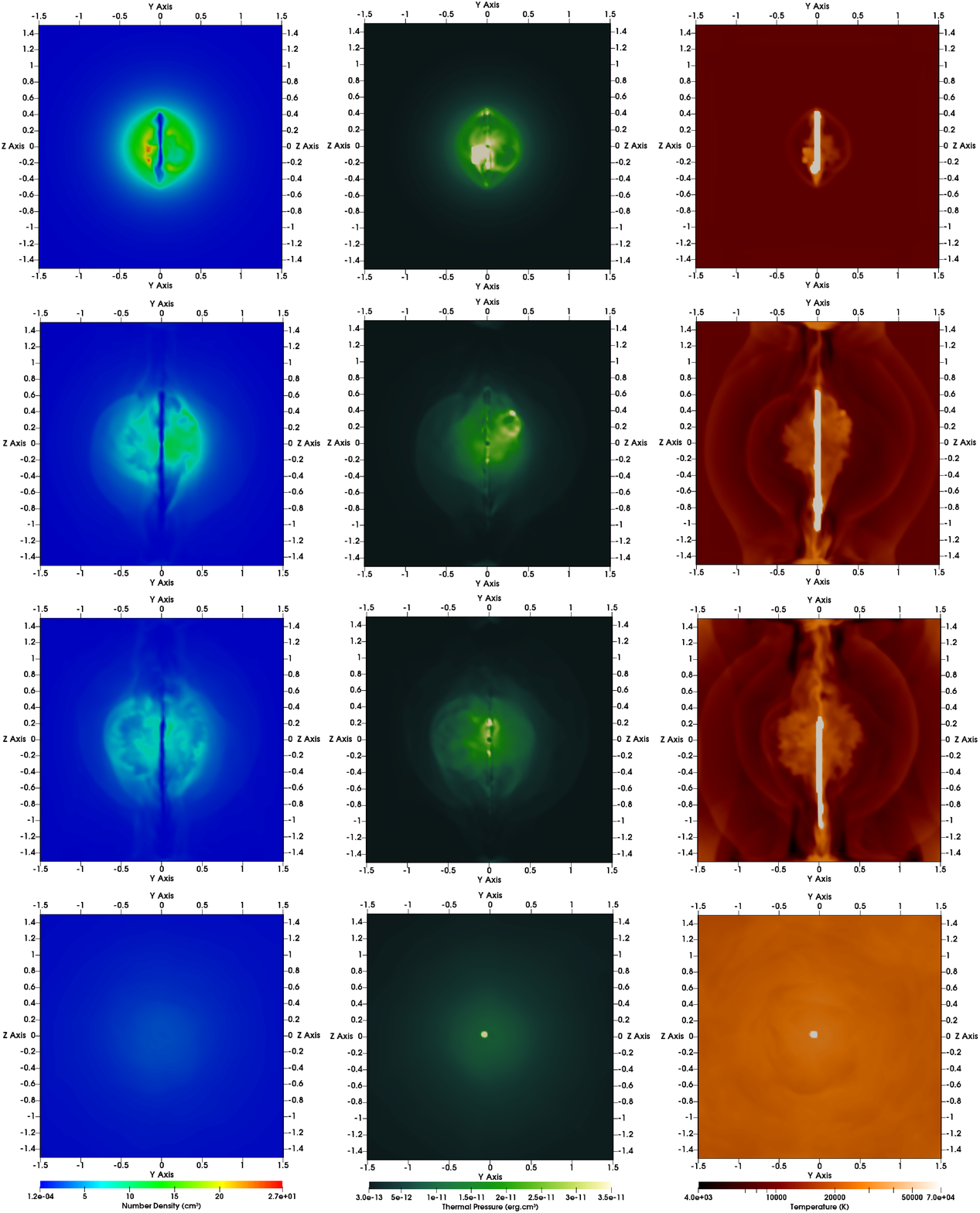}
\caption{Cut in the $x = 0$ plane of the galaxy at $t = 25$ Myr (upper line), $t = 100$ Myr (second line), $t = 150$ Myr (third line) and $t = 3$ Gyr (lower line), showing the gas density (left panel), thermal pressure (middle panel) and temperature (right panel) profiles for  Simulation JETSNE011000}
\label{dist_jetsne_150_011000_00530600}
\end{figure*}

A jet feature is created in the Simulations {JETSNE0031000, JETSNE0051000, and JETSNE011000} similar to the cases with an homogeneous medium. However, when SNe explode close to the axis of the initial velocity of the outflow, the motion of the gas influenced by the latter and the formation of the jet are affected (the interplay between the IMBH's outflow and the SNe feedback will be discussed further in details in a next paper). A feature similar to the one in the homogeneous medium, propagating almost freely and symmetric in both directions of the axis, can hardly be seen. Figure \ref{dist_jetsne_150_011000_00530600} exhibits the gas density, temperature and velocity profiles in a cut in the $x = 0$ plane of the galaxy at four different epochs, for Simulation JETSNE011000. On the top panel (25 Myr after the beginning of the simulation), the jet feature can be seen, but it is not linear nor symmetric. At 150 Myr (the third line in the Figure \ref{dist_jetsne_150_011000_00530600}), the superior part of the jet is briefly interrupted at $\sim$0.3 kpc, whereas the counter-jet remains unchanged, in the whole range from the center of the galaxy to its outskirts. This kind of interruption is repeated several times during the lifetime of the outflow, diminishing its effect in the gas removal of the galaxy. A comparison between two identical simulations, one with the SNe feedback and the IMBH's outflow (JETSNE011000) and other with only the SNE feedback, suggests that the amount of gas pushed away from the galactic center by the BH's outflow is negligible, when SNe feedback is also taken into account. 

\begin{figure}
 \centering
 \includegraphics[width=8cm]{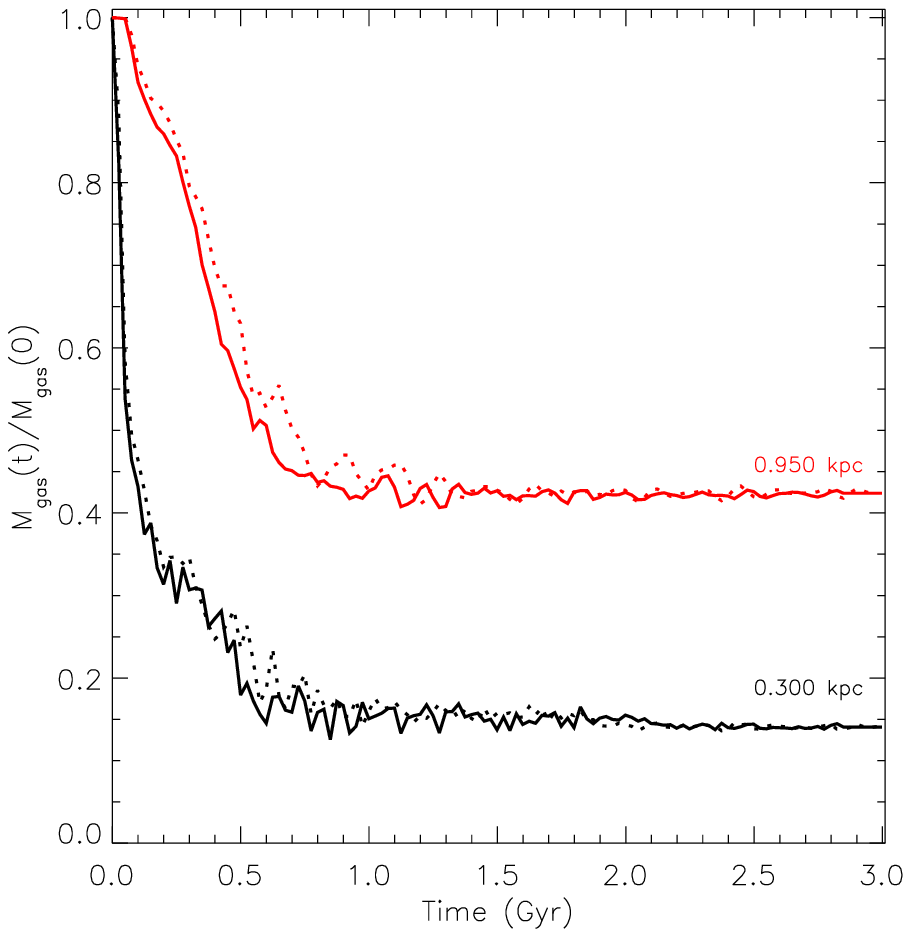}
\caption{The fraction of the initial mass as a function of time in two different galactic regions (300 pc - black lines - and 950 pc - red lines) for the simulation JETSNE011000 (thick solid lines) and a simulation without the BH outflow (thin dotted lines).}
 \label{gas_fraction_jetsne011000}
 \end{figure}

At the end of  3 Gyr of the simulation, the fraction of the initial mass remaining inside the galaxy at two different spherical regions in the simulation JETSNE011000 is virtually the same as in an identical simulation, but without the IMBH's outflow (Figure \ref{gas_fraction_jetsne011000}). The evolution in time of the remaining mass inside the galaxy is almost identical in both cases: an intense decrease in the amount of gas until approximately 600 Myr (when the peak of the SNe rate occurs) in the two regions and, after that, gas continues to leave each region but at a lower rate. At the end of 3 Gyr, the inner region (inside 300 pc) looses more gas, $85\%$ of the initial value, whereas at the tidal radius of the galaxy around $55\%$ of the gas is lost. This pattern, as pointed out in  \citet{cap15} and \citet{cap17} is totally consistent with the SNe rate (more intense in the first 600 Myr of the evolution of the galaxy and lower after that). The negligible contribution from the outflow to the gas removal, in fact, seems a bit surprising at first, especially due to the fact that in models without the SNe feedback the outflow is able to remove gas from the galaxy.

\section{Discussion} \label{sec:disc} 
 
From the results described in the previous section, two main facts emerge. The first one regards the parameter space study. There is a certain range of values for the initial density and initial speed of the IMBH's outflow in a dSph galaxy that allows it to impact the ISM and, depending on the scenario, create a jet feature. The speed parameter space is limited by kinematic measurements of extended outflows in isolated dwarf galaxies, possible associated with AGNs (\citealt{manza2019};\citealt{liu20}). These measurements point to outflow speeds as high as 1200 km s$^{-1}$ in galaxies with stellar mass in the range 10$^{8}$ to 10$^{9}$ M$_{\sun}$. In this sense, we adopted a maximum initial value of 2000 km s$^{-1}$, which decreases to below  a thousand km s$^{-1}$ as soon as the outflow starts interacting with the ISM. Since classical dSph galaxies are characterized by stellar masses below the range analysed by \citet{manza2019} it is reasonable to believe that the speeds of the gas driven by an outflow should be lower too. On the other hand, the influence of initial speeds below 1000 km s$^{-1}$ in the capability of the IMBH's outflow to flow through the ISM is negligible. In fact, decreasing this speed by a factor of 10-100 will not affect the gas along the way of the outflow's injection. The parameter space of the initial density of the outflow is constrained by the observationally estimated mass for IMBHs in the centers of dwarf galaxies - from  M $ \sim$ 10$^{4}$ to 10$^{6}$ M$_{\sun}$ yielding densities from 0.003 to 0.1 cm$^{-3}$, with the higher value being an upper limit already in the range of a massive black hole ($ 2 \times 10^{6} $ M$_\sun$), but still consistent with the value of $ 2.76 \times 10^{6} $ M$_\sun$ inferred for the black hole in Ursa Minor dSph galaxy by \cite{mann15}. When an IMBH in the lower mass range is considered, represented by the lowest initial densities, its outflow is not energetic enough to have an impact in the ISM, even when the density of the medium is almost homogeneous, without perturbations. However, if the central IMBH is in the mass range of $\sim 10^{5}$ to 10$^{6} $ M$_\sun$, the initial density of the outflow is higher and its energy is enough for it to break out into the medium, creating a jet feature and pushing gas to regions far from the tidal radius of the galaxy. 

The second fact to be considered is the lack of influence of the IMBH's outflow on the gas removal of a dSph galaxy when SNe feedback is also included into the simulations. 
When the galactic ISM is almost homogeneous, the IMBH's outflow alone can remove up to 10$\%$ of the initial gas of the galaxy at the end of 3 Gyr. However, when the energy injected by SNe in the ISM is also accounted for, the effects of the IMBH's outflow are reduced considerably: there is no significant difference in the amount of remaining gas mass inside the galaxy at the end of the simulation, compared to when only SNe feedback is considered. This result is in fully agreement with \citet{wagn12} who claimed that jet feedback becomes less efficient when $L_\mathrm{kin}/L_\mathrm{Edd} \la 10^{-4}$, where $L_\mathrm{Edd}$ is the Eddington luminosity. Indeed, $L_\mathrm{kin}/L_\mathrm{Edd}$ in our simulations ranges roughly from $10^{-11}$ to $10^{-4}$ considering the values of $L_\mathrm{kin}$ listed in Table \ref{table1} and the possible mass range for the IMBH in Ursa Minor ($\sim 10^4 - 10^6$ M$_{\sun}$; \citealt{macc05}; \citealt{lora09}; \citealt{nuci13}; \citealt{mann15}).
The possible explanation for this lies in the complexity of the internal galactic gas dynamics when both the SNe feedback and the IMBH's outflow are considered simultaneously. Previous works considered the energy or luminosity of the SNe and the BH feedback separately and compare them in order to evaluate the impact of both in the ISM (see for example \citealt{dashy17} and references therein). By doing this one is neglecting the interplay between such feedbacks (topic of a future work, already in preparation). Both objects, SNe and the IMBH, inject energy in the interstellar medium, but in different amounts, in different galactic locations, for different periods, in different ways. The energy of the SNe, for instance, is released everywhere in the galaxy, in all directions for a time interval of a few Gyr, whereas the outflow in our simulations has an initial velocity in just one direction for as long as some hundreds of  Myr. Because of this, the outflow, when energetic enough, creates a thin ``channel" through which the energy flows, without pushing ISM gas. Only the gas in front of the outflow is pushed away from the tidal radius of the galaxy, and just when it is being formed. Supernovae blasts, on the other hand, create large regions of high temperature and high pressure that could accelerate the gas, giving rise to galactic winds, that remove gas from the galaxy. Besides, Rayleigh-Taylor instabilities create hot "bubbles" of gas that could carry material away from the gravitational potential of the system (\citealt{ruiz13,cap15}). In larger galaxies, \citet{meli15} investigated the role of BH jets in active spiral galaxies and concluded similarly that jets alone do not impact the gas loss significantly, but can accelerate clumps of the underlying outflow to very high speed. On the other hand, \citet{bara19} performed cosmological simulations aiming to investigate the growth of IMBH in dwarf galaxies and their contribution to the gas removal and star formation quenching. They argued that black holes seeds can grow to a few $ M \sim 10^{5} - 10^{6} $ M$_\sun$ by $z\sim$ 5 and that they can decrease star formation by a factor of 3 when the IMBH has a mass a few times $10^{5} $ M$_\sun$. This late result is in agreement with our suggestion that higher density outflows impact more significantly the ISM of the galaxy, moving gas away. 

Strengthening the hypothesis that in dwarf galaxies BH feedback can be important in the gas removal and quenching of the SF, \citet{dashy17}, through an analytical approach, argued that there is a critical halo mass for dwarfs below which the feedback from a central black hole can remove gas from the galaxy and be more important than the stellar feedback. They suggested that AGN feedback can be an important driver of gas loss, more efficient than SNe. Their opposite conclusion, when compared to this work, could be related to the prescription for the energetics adopted by these authors: they considered a spherical BH outflow that impacts all the ISM of the galaxy and estimated the total luminosity due to the stellar feedback, whereas in this work the outflow affects just one axis of the galaxy, limiting its interaction with the interstellar medium. Besides, the SNe energy is injected into the ISM at each time-step following a specific SNe rate (constrained by the chemical properties of the galaxy) and the conditions described in section 3.3 for the location of the cell where the SNe energy is injected. 
As  described in previous sections, the SNe feedback and the ISM conditions can prevent the outflow from accelerating the gas, even though energy is being released into the medium. Kinetic luminosities from the BH outflow and from the SNe can compete with each other to accelerate the gas, depending how, where and when they are released in the medium. This could lead to very different results regarding the motion of gas throughout the galaxy.

The results discussed here regarding the contribution of the outflow to the mass loss could be viewed as a lower limit due to limitations in the prescription adopted for its occurrence. The outflow is kept unchanged during its lifetime, without interruptions or variations in its main parameters (density and velocity). If, for instance, an intermittent outflow is adopted (see \citealt{scha15}), turning on and off during its lifetime, more gas could be pushed away every time it is turned on again. When the outflow is turned off, the channel of low density that the jet feature creates can be filled with gas moved by supernovae, or even by gravity. This gas could then be accelerated and moved away from the galaxy when the outflow is turned on again. The same effect could result (and also be intensified) from an intermittent outflow with a varying initial velocity, if its direction is changed every time the outflow is turned on again. Another fact not taken into account in this work is the growth of the black hole and the varying mass accretion in time. These two facts could increase the outflow luminosity during its existence, increasing the impact on the medium. Besides that, the possible existence of overmassive black holes in dwarf galaxies is not considered in this work. \citet{koud20} investigated the impact of AGN feedback, particularly in dwarf galaxies, by means of cosmological simulations and found out that a fraction of dwarf galaxies exhibit large positive offsets from the $M_\mathrm{BH}-M_\mathrm{gas}$ or the $M_\mathrm{BH}-M_\mathrm{stellar}$ relations. In these galaxies with overmassive BHs, the outflow velocities and temperatures are higher, being able to escape the galaxy haloes and decreasing the gas reservoir. These scenarios are being analyzed and should be discussed in details in a future paper already in preparation.

\section{Conclusions} \label{sec:concl}

We have investigated whether and under what conditions the outflow of an IMBH can contribute to gas loss from a typical dSph galaxy. Non-cosmological 3D hydrodynamic simulations of an isolated typical dwarf spheroidal galaxy (using Ursa Minor as a template) taking into account  SNe feedback on the outflow of an IMBH were performed first to study the parameter space of the initial velocity and initial density of the outflow. After defining a range of values for these two parameters, the effects of the outflow in the ISM were analysed and the gas loss inferred. Two scenarios were considered in this work: an almost homogeneous medium, with the gas density varying only with  galactic radius, and another with the ISM disturbed by SNe explosions. 

Our main conclusions can be summarized as follows:

\begin{itemize}

 \item In both scenarios, with an almost homogeneous medium and with an ISM disturbed by SNe explosions, the outflow can only develop, impact the medium, and create a jet-like feature if its initial density and initial speed are in a range of values between 0.003 cm$^{-3}$ to 0.1 cm$^{-3}$ and 1000 km s$^{-1}$ to 2000 km s$^{-1}$, respectively. Values lower than these do not allow the outflow to leave the central cell and impact the ISM significantly. Higher values of density associated with BHs of masses larger than the ones inferred for this type of galaxy ($\sim 10^6$ M$_\sun$) and initial velocities higher than the upper limit are in contradiction with observations that estimate an upper value of $\sim 1200$ km s$^{-1}$ in outflows possibly associated with AGNs in dwarf galaxies;

 \item In the case of an almost homogeneous medium, the outflow creates a jet-like feature that propagates freely and symmetrically in both directions of the velocity axis with the same speed (lower than the initial value). The outflow removes a small fraction of the gas from the galaxy (the fraction depends on the initial physical conditions of the outflow); 

 \item When inhomogeneities created by SNe are taken into account, the effects of the outflow are substantially reduced. A jet-like feature is also created, but the symmetry can be broken by turbulence in the medium due to the multiple SNe explosions. Depending on the initial conditions of the outflow, the jet-like feature can be even destroyed. The speed of the propagation in the jet is also lower than in the case of a homogeneous medium;

 \item Outflows with higher initial densities and initial speeds have a greater impact in the ISM, whereas in the case of lower values, only the region around the velocity axis is affected. The amount of gas removed due to the outflow is also larger in the former case;

 \item The outflow alone, when SNe are not taken into account, can remove up to 10$\%$ of the gas of the ISM. However, if SNe are also considered, the contribution of the outflow to the gas removal becomes negligible. Two models, one with only SNe feedback and another with SNe and the IMBH's outflow, exhibit the same final gas mass inside two different galactic radii (300 pc and 950 pc);
 
 \item In the most general scenario adopted in our simulations (IMBH outflow with initial density and initial velocity within the adopted range, leaving the galactic center in just one direction for a continuous time interval in an ISM disturbed by SNe explosions), the contribution of the mechanical feedback of an IMBH to the gas loss is negligible when compared to the SNe feedback;
 
 \item The interplay between the SNe feedback and the IMBH outflow could explain the lack of influence of the outflow on  gas loss in dwarf spheroidal galaxies.

\end{itemize}

\acknowledgments

The authors thank FAPESP grants 2017/25779-2, 2017/03173-4, 2017/25651-5, and 2014/11156-4 and acknowledge the National Laboratory for Scientific Computing (LNCC/MCTI, Brazil) for providing HPC resources of the SDumont supercomputer. URL: http://sdumont.lncc.br. 
 
\appendix


\section{Influence of the resolution: a convergence study}

\begin{figure*}
 \centering
 \includegraphics[width=14cm]{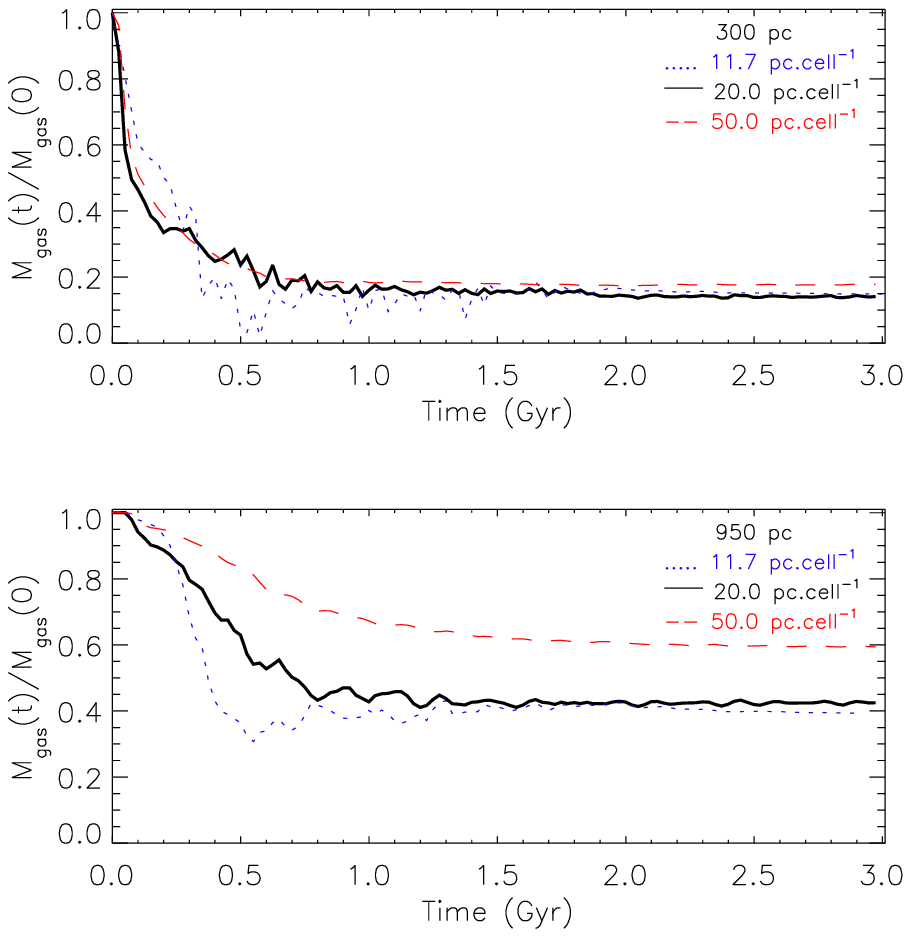}
 \caption{The fraction of the initial mass as a function of time inside 300 pc radius (above) and 950 pc radius (below) in three identical simulations, but with different resolutions: 11.7 pc.cell$^{-1}$ from \citet{cap17} (blue short-dashed line), 20.0 pc.cell$^{-1}$ from this work (black straight line) and a test model with 50.0 pc.cell$^{-1}$ (red long-dashed line).}  
 \label{fig10}
 \end{figure*}

	In order to investigate how the resolution adopted in the simulations could affect the results here presented, we compare the fraction of the initial gas that is lost in three simulations with different number of cells in computational boxes with the same size. The initial setup and all other configurations are the same in all cases. In all simulations in this work, a computational box of 3.0 kpc side was adopted, divided in 150 cell each side, resulting in a resolution of 20.0 pc cell$^{-1}$, lower than the resolution adopted in the first two papers of the group (11.7 pc cell$^{-1}$) (\citealt{cap15}; \citealt{cap17}). We also ran a test simulation with 60 cells in each side of the box, yielding a resolution of 50.0 pc cell$^{-1}$. In Figure \ref{fig10}, it is shown the fraction of the initial mass as a function of time for the three simulations, inside two different regions: the 300 pc central radius (upper panel) and 950 pc - approximately the tidal radius of the model galaxy, Ursa Minor dSph (bottom panel). In the central 300 pc, where the BH's outflow is more effective, the residual mass is essentially the same during the whole period simulated for the three resolutions, even the lowest one. There is a fast decrease in the gas mass in the first 500 to 600 Myr, and a plateau around 15$\%$ of the initial mass until the end in all cases. During the decrease, there are small differences in the fraction of the initial mass which, however, do not affect the general results. These differences vanish after $\sim$ 1 Gyr, specially in the cases with 11.7 and 20.0 pc cell$^{-1}$ (blue and black lines respectively). When the lowest resolution (50.0 pc cell$^{-1}$) is considered, the final gas mass is slighty higher (around 18$\%$ of the initial mass).

	If the mass fraction is computed inside a larger region, up to 950 pc, the resolution becomes more important and significant differences appear. The gas loss in the case with the lowest resolution is also lower, by a factor of $\sim 1.5$ compared to the other two cases. At the end of the 3 Gyr, the simulations with 11.7 and 20.0 pc cell$^{-1}$ lose almost 60$\%$ of the initial mass whereas 40$\%$ is lost when the resolution is 50.0 pc cell$^{-1}$. There is also a difference in the fast decrease of the gas mass between the simulations with 11.7 and 20.0 pc cell$^{-1}$: a lower resolution implies a lower decrease. After approximately 1 Gyr and up to the end, however, the amount of gas that is lost is the same. 
	
	Since the outflow acts mainly in the central region of the galaxy and considering the very small differences in the gas loss (inside 300 pc or 950 pc) when 11.7 or 20.0 pc cell$^{-1}$ resolutions are adopted, we believe that is safe to argue that our results can be considered stable with the adopted resolution of 20.0 pc cell$^{-1}$. Resolutions lower than that, such as the one tested here, could affect results regarding the gas loss inside the whole galaxy. However, it should also be strengthened that the main goal of the paper is not to infer the amount of mass that is lost, but rather analyze the influence of the outflow on the mass loss.

\end{document}